\documentclass{article}

\usepackage{authblk}
\usepackage{amsmath}
\usepackage{graphicx}
\usepackage[font=small]{caption}

\usepackage{setspace}
\let\citeleft=(
\let\citeright=)

\begin{document}

\pdfinfo{
   /Author (AUTHORS)
   /Title (TITLE)
}

\title{\vspace{-2cm}Accelerating cardiac cine MRI using a deep learning-based ESPIRiT reconstruction}

\author[1]{Christopher M. Sandino}
\author[2]{Peng Lai}
\author[3]{Shreyas S. Vasanawala}
\author[3]{Joseph Y. Cheng}

\affil[1]{\small Department of Electrical Engineering, Stanford University, Stanford, CA, 94305, United States}
\affil[2]{\small Applied Sciences Laboratory, GE Healthcare, Menlo Park, CA, 94025, United States}
\affil[3]{\small Department of Radiology, Stanford University, Stanford, CA, 94305, United States}

\date{}
\maketitle

\vfill
\noindent
\textit{Running head:} DL-ESPIRiT

\noindent
\textit{Address correspondence to:} \\
  Christopher M. Sandino \\
  Department of Electrical Engineering \\
  Stanford University, Stanford, CA, 94025, United States \\
  sandino@stanford.edu

\noindent
This work was supported by NSF Graduate Research Fellowship, General Electric Healthcare, Google Cloud.

\noindent
Approximate word count: 249 (Abstract) 5000 (body)\\

\noindent
Submitted to \textit{Magnetic Resonance in Medicine} as a Full Paper.

\clearpage

\section*{Abstract}

\noindent
\textbf{Purpose}:  To propose a novel combined parallel imaging and deep learning-based reconstruction framework for robust reconstruction of highly accelerated 2D cardiac cine MRI data. 

\noindent
\textbf{Methods}: We propose DL-ESPIRiT, an unrolled neural network architecture that utilizes an extended coil sensitivity model to address SENSE-related field-of-view (FOV) limitations in previously proposed deep learning-based reconstruction frameworks. Additionally, we propose a novel neural network design based on (2+1)D spatiotemporal convolutions to produce more accurate dynamic MRI reconstructions than conventional 3D convolutions. The network is trained on fully-sampled 2D cardiac cine datasets collected from eleven healthy volunteers with IRB approval. DL-ESPIRiT is compared against a state-of-the-art parallel imaging and compressed sensing method known as $l_1$-ESPIRiT. The reconstruction accuracy of both methods is evaluated on retrospectively undersampled datasets (R=12) with respect to standard image quality metrics as well as automatic deep learning-based segmentations of left ventricular volumes. Feasibility of DL-ESPIRiT is demonstrated on two prospectively undersampled datasets acquired in a single heartbeat per slice.

\noindent
\textbf{Results}: The (2+1)D DL-ESPIRiT method produces higher fidelity image reconstructions when compared to $l_1$-ESPIRiT reconstructions with respect to standard image quality metrics ($P<$0.001). As a result of improved image quality, segmentations made from (2+1)D DL-ESPIRiT images are also more accurate than segmentations from $l_1$-ESPIRiT images. 

\noindent
\textbf{Conclusion}: DL-ESPIRiT synergistically combines a robust parallel imaging model and deep learning-based priors to produce high-fidelity reconstructions of retrospectively undersampled 2D cardiac cine data acquired with reduced FOV. Although a proof-of-concept is shown, further experiments are necessary to determine the efficacy of DL-ESPIRiT in prospectively undersampled data.
  
\noindent
\textbf{Keywords}: cardiac cine, compressed sensing, deep learning

\clearpage

\section{Introduction}
\label{sec:introduction}
Cardiac cine MRI is a widely used imaging technique for non-invasive characterization of heart morphology and function \cite{Sakuma1993}. In a standard cardiac cine MRI scan, a two-dimensional (2D) steady-state gradient echo acquisition is synchronized with the cardiac cycle and typically performed over 10-15 slices covering the entire heart in a short-axis view. To minimize motion artifacts caused by respiration, the patient is asked to repeatedly hold their breath for 15-20 seconds at a time until all slices are acquired. 

Such multi-breath-hold protocols pose several problems to the cardiac MRI workflow. Firstly, exams are long and complex due to the need of repeated breath-holds and intermediate resting periods for patient recovery. Secondly, long successive breath-holding is uncomfortable, and can be especially challenging in patients with impaired breath-hold capacity. Finally, the inevitable variation in the level of inspiration across each breath-hold introduces slice misalignment in the cine images. This variation is known to impact the ability to accurately estimate ventricular volumes \cite{Greil2008}, which are used to compute important functional indices such as left ventricular ejection fraction (LVEF). These issues can be mitigated by reducing the total scan time and thereby reducing the number of breath-holds. In fully-sampled imaging, the only way to achieve this is to trade-off in-plane spatial resolution, number of slices, and/or temporal resolution, each of which may impact diagnostic value. Reductions in scan time can also be achieved by using more efficient non-Cartesian sampling trajectories to traverse k-space \cite{Roifman2019}. In this work, we only consider the standard Cartesian sampling since non-Cartesian acquisitions remain susceptible to various image artifacts related to off-resonance and system imperfections. 

Scan time can also be reduced by parallel imaging (PI) methods that leverage multi-channel receiver array information to reconstruct the image from data sampled below the Nyquist criteria. For example, sensitivity encoding (SENSE) utilizes explicit knowledge of coil sensitivity profiles to localize signals in space and remove aliasing artifacts introduced by undersampling \cite{Pruessmann1999}. However, SENSE-based methods require accurate estimation of coil sensitivity maps from calibration data; otherwise, model errors can arise resulting in reconstruction artifacts. For example, when the field of view (FOV) is smaller than the subject, overlapping anatomies create discontinuities in coil sensitivity maps and cause residual ghosting in the SENSE reconstructed images \cite{Griswold2004,Goldfarb2004}. When imaging at double oblique scan planes, such as standard cardiac cine views, the FOV must be prescribed conservatively large to avoid such errors. k-Space based PI approaches, such as GRAPPA \cite{Griswold2002} or SPIRiT \cite{Lustig2010}, do not rely on explicit sensitivity maps and instead exploit coil-wise correlations in k-space to directly synthesize missing data samples. Generally, k-space based approaches are robust to anatomy overlap and can enable faster scans with a reduced FOV \cite{Blaimer2004}. Either type of method can be used to accelerate scan time by a factor of 2-3X without sacrificing image quality or resolution. For this reason, parallel imaging is almost always used in routine clinical scans to reduce the number of breath-holds; however, 5-6 breath-holds are still necessary for a standard 2D cardiac cine acquisition of 10-12 slices \cite{Kramer2013}. 

Further scan time acceleration has been achieved by exploiting prior information about the underlying signal structure in addition to parallel imaging during reconstruction. Many reconstruction methods have been developed to leverage spatiotemporal redundancy in dynamic imaging data to remove aliasing artifacts \cite{Kellman2001,Breuer2005,Tsao2003}. Other methods, such as compressed sensing (CS), instead leverage incoherent data sampling and spatiotemporal transform sparsity to reconstruct highly undersampled k-space data by iteratively solving a regularized inverse problem \cite{Lustig2007,Jung2009,Feng2013}. CS methods have been instrumental in enabling rapid 2D cardiac cine acquisitions that can be completed in a single breath-hold \cite{Vincenti2014,Kido2016} or while the patient is freely breathing \cite{Feng2013,Xue2013}. Despite its potential to drastically simplify the cardiac MRI exam workflow, CS suffers from multiple challenges which limit the maximum achievable scan time acceleration. Transform sparsity assumptions used by CS are often simplistic and incapable of accurately modelling complex cardiac dynamics. Thus, CS requires careful hand-tuning of the relative weights assigned to data consistency and regularization terms. However, reconstruction from highly undersampled data requires strong regularization to completely remove aliasing artifacts, which leads to over-regularization that produces images with textural artifacts and spatiotemporal blurring. 

More recently, deep learning-based approaches have been proposed to leverage historical exam data from multiple subjects to implicitly and automatically learn better priors for constrained image reconstruction \cite{Yang2016,Hammernik2018,Cheng2018,Aggarwal2019,Sandino2020}. These methods are comprised of the following steps: 1) a conventional CS algorithm is unrolled to a fixed number of iterations, 2) the prior information in each iteration is enforced by a neural network, and 3) the unrolled algorithm is trained end-to-end in a supervised fashion. For 2D cardiac cine imaging, various neural network architectures, including cascaded \cite{Schlemper2018} and recurrent unrolled networks \cite{Qin2019,Qin2019a}, have been developed and trained to reconstruct up to 9X accelerated data with higher fidelity than CS. However, Refs.~\cite{Schlemper2018,Qin2019,Qin2019a} did not leverage parallel imaging information which could potentially enable further scan time acceleration. Several other deep learning-based reconstruction methods that do leverage parallel imaging \cite{Hammernik2018,Cheng2018} use a limited coil sensitivity model that remains susceptible to SENSE-related FOV limitations. 

In this work, a combined parallel imaging and deep learning-based reconstruction framework is proposed for robust reconstruction of dynamic MRI data. The novelty of this framework is summarized as follows: 1) an extended coil sensitivity model based on ESPIRiT \cite{Uecker2014} is integrated into an existing deep learning-based reconstruction approach \cite{Sandino2020} to improve its robustness to SENSE-related FOV limitations and 2) a novel convolutional neural network (CNN) architecture based on separable 3D convolutions is developed to learn spatiotemporal priors for dynamic data more efficiently. We apply our novel DL-ESPIRiT network to 12X retrospectively undersampled 2D cardiac cine data and show higher reconstruction accuracy than a state-of-the-art parallel imaging and compressed sensing (PICS) algorithm with respect to standard image quality metrics. Furthermore, we show that as a result of improved image quality and sharpness, the accuracy of automatic LVEF measurements is improved over the PICS reconstruction method. Feasibility of this approach is demonstrated in two prospectively undersampled datasets which are acquired at a rate of one single heartbeat per slice.

\section{Theory}
\label{sec:theory}
\subsection{Reconstruction Overview}
The MR imaging process can be modelled as a linear system of the form:
\begin{align} \label{eq:sense}
    y &= PFSx \\
    S &= [S^1, ..., S^N]^T
\end{align}
where the true image ($x$) is transformed into the sampled multi-channel raw data ($y$) using a forward model comprised of the coil sensitivity operator ($S$), discrete Fourier transform ($F$), and k-space sampling operator ($P$). If all $N$ coil sensitivity maps ($S^i$) can be reliably estimated from calibration data, then a SENSE \cite{Pruessmann1999} parallel imaging reconstruction $\hat{x}$ can be directly obtained from $y$ using the least squares solution to Eq.~\ref{eq:sense}.

Alternatively, ESPIRiT \cite{Uecker2014} exploits advantages of k-space based PI techniques by deriving SENSE-like sensitivity maps using an eigenvalue approach. Multiple sets of ESPIRiT maps can be derived from calibration data to flexibly represent overlapping anatomies in reduced FOV acquisitions using an augmented forward model:
\begin{align} \label{eq:espirit}
    y &= PF \sum_{i=1}^M S_i x_i \\
    S_i &= [S^1_i, \dots, S^N_i]^T
\end{align}
where $M$ is the number of sets of ESPIRiT maps, $S_i$ is the $i$-th set of ESPIRiT maps, and $x_i$ is the corresponding $i$-th image. The number of sets of ESPIRiT maps (M) is a hyperparameter that must be chosen prior to reconstruction. As suggested by Ref.~\cite{Uecker2014}, two sets of maps in practice are sufficient for high-fidelity reconstruction of images with anatomy overlap. For simplicity, Eq.~\ref{eq:espirit} can be re-written in the same form as Eq.~\ref{eq:sense} by vertically stacking images $x_i$ and sensitivity map operators $S_i$:
\begin{align} \label{eq:espirit2}
    y &= PFEx \\
    E &=
    \begin{bmatrix}
    S_1^1 & \dots & S_M^1 \\
    \vdots & \ddots & \vdots \\
    S_1^N & \dots & S_M^N
    \end{bmatrix} \\
    x &= [x_1, \dots, x_M]^T
\end{align}
into a single ESPIRiT matrix $E$. Here, the $P$ and $F$ operators are repeated for each coil image. Similarly to SENSE, multiple ESPIRiT reconstructions $\hat{x_i}$ can be obtained from y using the least squares solution of Eq. \ref{eq:espirit2}.

For datasets acquired with an acceleration rate that is higher than what is supported by the coil hardware, the SENSE and ESPIRiT problems become ill-posed leading to noise amplification and residual aliasing artifacts in the reconstructed images. These can be suppressed by incorporating prior information about the images into the ESPIRiT reconstruction via regularization. The regularized ESPIRiT problem can be formulated as a non-linear inverse problem of the form:
\begin{equation}\label{eq:regls}
    \hat{x} = \underset{x}{\text{arg min }}  || y - PFEx ||_2^2 + \lambda R(x)
\end{equation}
where $R$ is a regularization function with associated regularization strength $\lambda$. In compressed sensing theory, the sampling operator (P) is designed to pseudo-randomly sample k-t space causing aliasing artifacts in image domain to appear incoherent and noise-like \cite{Lustig2007}. Thus, the regularization term is typically designed to suppress noise-like artifacts in the final images by promoting sparsity in some transform domain, such as in $l_1$-ESPIRiT. More generally, when $R$ is a proper convex function, the proximal gradient descent (PGD) method \cite{Parikh2014} can be used to iteratively solve the optimization problem in Eq.~\ref{eq:regls} by alternating between two updates. The first is a data consistency update of the form: 
\begin{equation}\label{eq:pgd1}
    \hat{x}^{k+1/2} = \hat{x}^k - 2t A^H (y - A \hat{x}^k)
\end{equation}
where $A$ is the forward ESPIRiT signal model ($PFE$), $A^H$ is its conjugate transpose, and $t$ is the PGD step size. This is followed by a proximal update of the form:
\begin{equation}\label{eq:pgd2}
    \hat{x}^{k+1} = P_{\lambda R}(\hat{x}^{k+1/2})
\end{equation}
where $P_{\lambda R}$ is the proximal operator of $\lambda R$ defined as:
\begin{equation}\label{eq:prox}
    P_{\lambda R}(v) = \underset{u}{\text{arg min }} R(u) + \frac{1}{2 \lambda} ||u - v||_2^2.
\end{equation}
In the case that $R$ is the $l_1$-norm of a unitary transform ($\Psi$) applied to $x$, i.e. $R(x) = \sum_{i=1}^M ||\Psi(x_i)||_1$, the proximal operator simplifies into a soft-thresholding function in the transform domain.

\subsection{Data-driven Reconstruction Overview}
Hand-crafted regularization functions have enabled significant acceleration of standard cardiac cine scans. However, data-driven regularization design can yield regularization functions that more accurately describe complex signal dynamics, and produce higher fidelity reconstructions as a result. In a deep-learning-based data-driven reconstruction, the regularization function $R$ is parameterized by a neural network, and automatically learned from historical exam data via some training process.

While other approaches directly learn $R$ using a field-of-experts model \cite{Hammernik2018}, a simpler and more straightforward approach is to learn the proximal operator of $R$ \cite{Diamond2017,Mardani2018}. This can be achieved by unrolling \cite{Gregor2010} the PGD algorithm in Eqs.~\ref{eq:pgd1} and \ref{eq:pgd2} to a fixed number of iterations, and replacing the regularization function's proximal operator with a neural network. This produces an unrolled PGD network of the form:
\begin{align}\label{eq:deep_pgd}
    \hat{x}^{k+1} &= G^k \big( \hat{x}^k - 2t A^H (y - A \hat{x}^k); \theta^k \big) \\
    k &= 1, \dots, K
\end{align}
where $G^k$ is a conditionally generative neural network whose parameters $\theta^k$ are learned uniquely for each of the $K$ iterations. The unrolled PGD network is trained end-to-end in a supervised fashion to output images which are close to fully-sampled reference images. Closeness is evaluated with respect to some dissimilarity metric $D$, for which common choices include pixel-wise $l_1$ and $l_2$ differences. A loss function $L$ is constructed from the average dissimilarity over all of the training examples:
\begin{equation}\label{eq:loss}
    L = \frac{1}{N_{\text{train}}} \sum_{i=1}^{N_{\text{train}}} D(\hat{x}_i^K, x_i)
\end{equation}
where $N_{\text{train}}$ is the number of training examples indexed by $i$, $\hat{x}_i^K$ is the unrolled PGD network output, and $x_i$ is the fully-sampled image. The loss function is iteratively minimized over the training dataset with respect to the network parameters $\theta^k$ by a stochastic gradient descent algorithm. 

\section{Methods}
\subsection{Network Architecture}
\begin{figure}[t]
\centering
\includegraphics[width=1.0\textwidth]{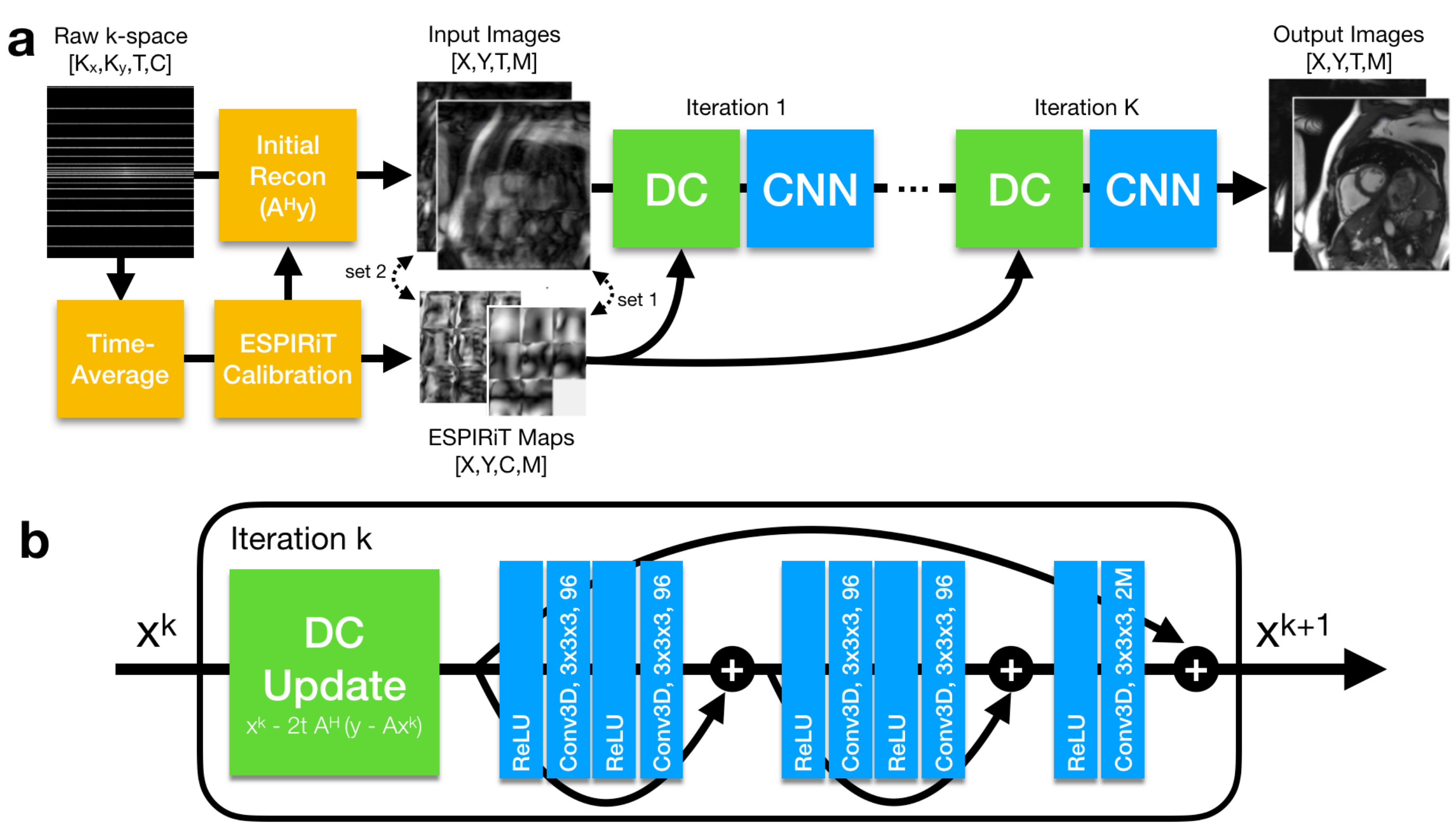}
\caption{(a) The full DL-ESPIRiT reconstruction pipeline is shown here. A fully-sampled calibration region is extracted from time-averaged k-space data and used to compute ESPIRiT maps. These maps are used to compute an initial zero-filled reconstruction from the raw k-space data with M complex-valued channels, where M is the number of sets of ESPIRiT maps computed during the calibration process. (b) The network iteratively applies data consistency (DC) updates, as described by Eq.~\ref{eq:pgd1}, followed by 3D CNNs. The 3D kernel size and the number of output channels of each convolutional layer are denoted inside each ``Conv3D'' block.}
\label{fig:network}
\end{figure}

The DL-ESPIRiT network architecture takes directly after the unrolled proximal gradient descent network described in the previous section. As shown in Figure \ref{fig:network}, the DL-ESPIRiT network takes inputs of a zero-filled reconstruction of a 2D cardiac cine slice, as well as its corresponding ESPIRiT maps computed from time-averaged k-space data. The network then alternates between 3D spatiotemporal convolutional neural networks (CNN) and data consistency steps. It is trained to output images which are close to the corresponding fully-sampled ground truth images in an $l_1$ pixel-wise sense:

\begin{equation}\label{eq:l1loss}
    L = \frac{1}{N_{\text{train}}} \sum_{i=1}^{N_{\text{train}}} ||\hat{x}_i^K - x_i||_1.
\end{equation}

Much like in the PGD algorithm, the data consistency step applies a gradient descent update in which the gradient of the data consistency term in Eq. \ref{eq:regls} is subtracted from the output of the CNN. In this step, ESPIRiT maps are used by the signal model to project data back and forth between image and k-space domains. This step ensures that the final output of the DL-ESPIRiT network does not deviate from the acquired k-space data.

In between data consistency and CNN updates, the complex-valued data is converted into two real-valued channels by stacking the real and imaginary parts along the channel dimension. Additionally, images corresponding to each set of ESPIRiT maps are also stacked along the channel dimension. For example, when using two sets of ESPIRiT maps, a total of four image channels is passed as input to the CNN. This way information across multiple images is shared allowing the CNN to learn a joint regularization function across the ESPIRiT channel dimension. In contrast, $l_1$-ESPIRiT regularizes each channel independently since it is not obvious how to jointly regularize these images. At the end of each CNN update, the real-valued image channels are converted back into complex-valued images in preparation for the subsequent data consistency update.

Each neural network between data consistency steps is a fully convolutional residual network (ResNet), which is currently the state-of-the-art architecture for computer vision tasks \cite{He2016}. Each ResNet is composed of 3D convolutional layers with $3 \times 3 \times 3$ kernels to leverage all spatial and temporal dimensions for de-aliasing. Convolutional layers are implemented using circular padding along the phase encoding and temporal directions in order to enforce circular boundary conditions in the two dimensions \cite{Sandino2020}. There are five total convolutional layers for each DL-ESPIRiT iteration which corresponds to a spatiotemporal receptive field of size $11 \times 11 \times 11$. The first convolution of each ResNet expands the initial (2 or 4) images into 96 feature maps, which are propagated through the network until the final convolution where they are recombined into the original number of input images. All convolutional layers are preceded by ReLU pre-activation layers, which operate separately on each image/feature channel \cite{He2016a}. Batch normalization layers are omitted from the network to reduce GPU memory costs during training.

\begin{figure}[t]
\centering
\includegraphics[width=0.8\textwidth]{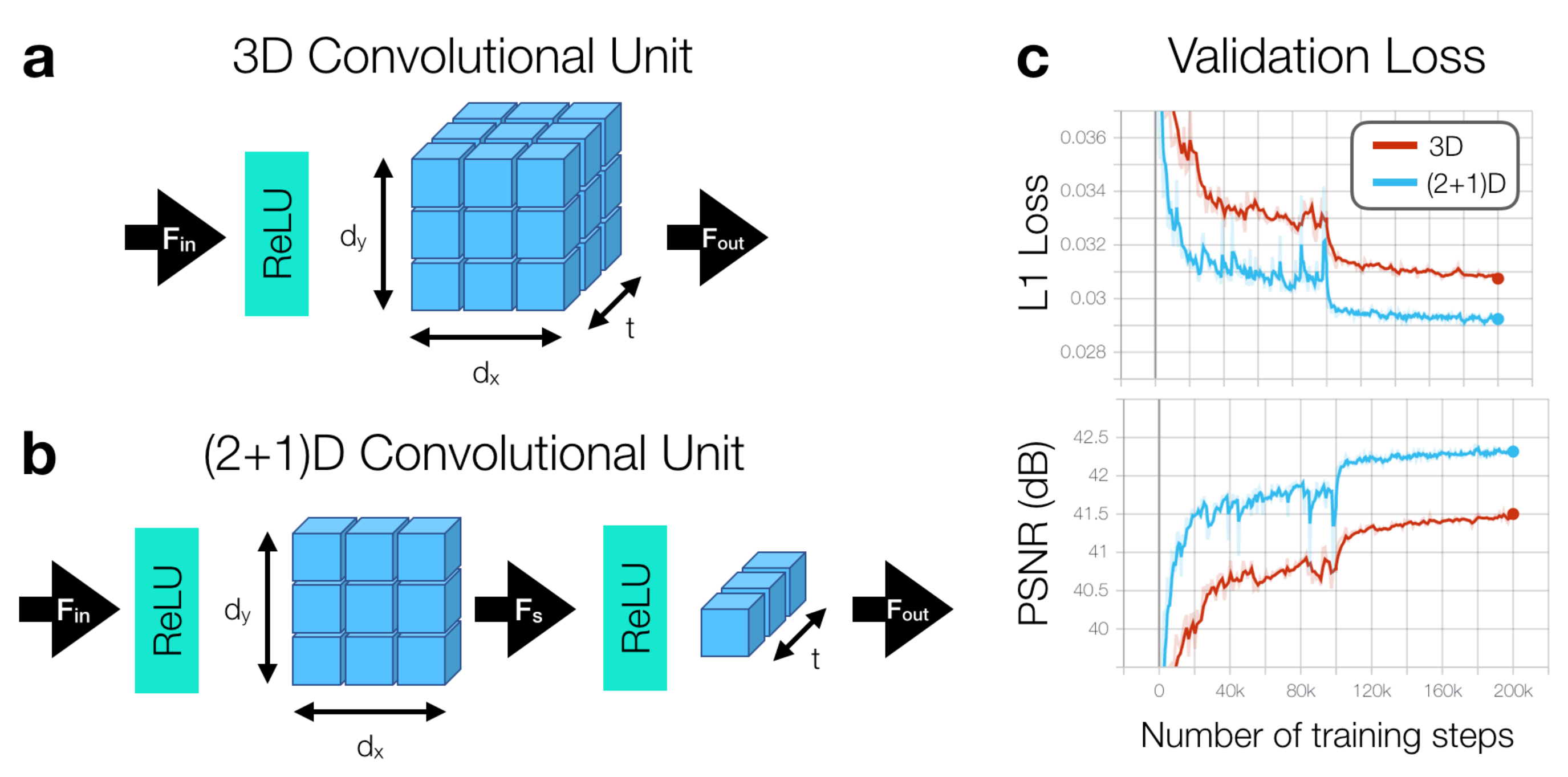}
\caption{(a) A 3D convolutional unit is composed of a ReLU pre-activation followed by a 3D convolution applied through 2D space and time that transforms  $F_{\text{in}}$ input channels into  $F_{\text{out}}$ output channels. (b) A separable 3D convolution, or (2+1)D convolution, is decomposed into simpler 2D spatial and 1D temporal convolutional layers with an additional ReLU activation in between.  (c) Validation loss curves are shown throughout training for 3D and (2+1)D reconstruction networks. Through the decomposition process, training convergence is relaxed allowing for lower validation loss and higher peak signal-to-noise ratio (PSNR) across the validation datasets. A warm restart with a smaller learning rate is performed after 100k training steps to further improve training convergence.}
\label{fig:separable}
\end{figure}

Inspired by previous work in CNN architecture design for video-based action recognition tasks, we also implement the DL-ESPIRiT network using separable 3D convolutions, henceforth referred to as (2+1)D convolutions \cite{Tran2018}. In this version of the network, all 3D convolutions are replaced with (2+1)D convolutions which have been decomposed into simpler 2D spatial and 1D temporal components (Fig.~\ref{fig:separable}B). An additional ReLU activation is added in between 2D spatial and 1D temporal convolutions to enhance the network's representational power. As shown by Figure \ref{fig:separable}C, (2+1)D convolutional kernels are simpler to learn resulting in lower training loss and higher reconstruction accuracy \cite{Sandino2018}. To make a fairer comparison, the number of learnable parameters between 3D and (2+1D) convolutions is matched by expanding the number of 2D spatial filters to be:
\begin{equation}\label{eq:separable}
    F_{\text{s}} = \Bigg\lfloor \frac{t d_x d_y F_{\text{in}} F_{\text{out}}}{d_x d_y F_{\text{in}} + t F_{\text{out}}} \Bigg\rfloor
\end{equation}
where the spatial kernel size is $d_x \times d_y \times 1$, the temporal kernel size is $1 \times 1 \times t$, $F_\text{in}$ is the number of input feature maps, $F_\text{out}$ is the number of feature maps output by the temporal convolution, and $F_\text{s}$ is the number of features output by the intermediate spatial convolution.

\subsection{Data Acquisition}
Fully-sampled, multi-slice 2D cardiac cine datasets were collected from 22 healthy volunteers at various scan planes using an ECG-gated balanced steady-state free precession (bSSFP) sequence with IRB approval. Each volunteer was asked to repeatedly hold his or her breath for 15-20 seconds during the acquisition of each slice. Every breath-hold was followed by adequate resting time to allow the subject to recover and ensure consistent breath-holding. Data collection was performed on a mixture of 1.5T and 3.0T GE MRI scanners (GE Healthcare, Waukesha, WI) using a 32-channel cardiac coil. To reduce the appearance of banding artifacts in the heart tissue, shim boxes were carefully placed on a tight region-of-interest around the heart prior to scanning. The relevant scan parameters are shown in Table \ref{tab:params}. For each volunteer, data were acquired at multiple slices and at different cardiac views including standard short-axis and long-axis (2-chamber, 3-chamber and 4-chamber) views. All datasets were coil compressed \cite{Zhang2013} from 32 to 8 virtual coils for computational speed and memory considerations. 

To prove the feasibility of the proposed DL-ESPIRiT reconstruction method, a customized ECG-gated bSSFP sequence is implemented to support prospective acquisition of each slice in a single heartbeat. Data acquisition of each slice is preceded by a whole cardiac cycle of dummy pulses to establish signal steady state. The k-t view-ordering is designed using a variable-density undersampling mask similar to the masks used for training \cite{Lai2014}. To reduce the appearance of banding artifacts and improve temporal resolution, the repetition time is minimized using a partial echo acquisition. With IRB approval, two pediatric patients referred for a routine cardiac MRI exam are scanned using the clinical standard cine sequence with SENSE acceleration (R=2), and the custom sequence with variable-density k-t acceleration (R=12). The relevant scan parameters are also shown in Table \ref{tab:params}.

\begin{table} 
\centering
\begin{tabular}{|l||c||c|c||c|c|}
\hline
 & \textbf{Volunteers} & \multicolumn{2}{|c||}{\textbf{Patient 1 (Fig.~4)}} & \multicolumn{2}{|c|}{\textbf{Patient 2 (Fig.~9)}} \\
\hline
& R=1 & R=2 & R=12 & R=2 & R=12\\
\hline\hline
Echo Time (ms) & 1.6 & 1.5 & 1.2 (partial) & 1.5 & 1.3 (partial)\\
Repetition Time (ms) & 3.7 & 3.5 & 3.4 & 3.5 & 3.5\\
Flip Angle ($^{\circ}$) & 40-60 & 50 & 50 & 50 & 50\\
Number of Slices$^*$ & 1-12 & 14 & 14 & 18 & 18\\
Slice Thickness (mm) & 8 & 8 & 8 & 8 & 8\\
Slice Gap (mm) & 2 & 0 & 0 & 0 & 0\\
Matrix Size & 200-224 x 160-180 & 200 x 160 & 200 x 144 & 200 x 170 & 200 x 170\\
Spatial Resolution (mm$^2$) & 1.8-2.0 x 1.6-1.8 & 1.5 x 1.5 & 1.5 x 1.5 & 1.5 x 1.5 & 1.5 x 1.5\\
Temporal Resolution (ms) & 40-42 & 44 & 44 & 53 & 53\\ 
Acquired Cardiac Phases & 20-25 & 12 & 12 & 25 & 25\\
Heart Rate (bpm) & 50-70 & 112 & 112 & 45 & 45\\ 
Scan Time$^{**}$ (s) & 15-250 & 96 & 17 & 283 & 48\\
\hline
\end{tabular}
\caption{Table containing scan parameters for fully-sampled training and prospective data collection. 
All training datasets were acquired from 22 healthy volunteers using a fully-sampled cine acquisition (R=1). Two datasets were acquired from each pediatric patient with 2-fold uniformly undersampled cine acquisition (R=2), and 12-fold k-t undersampled cine acquisition (R=12). $^{*}$For fully-sampled training data, the number of slices varied depending on the scan plane. Ten to twelve slices were acquired in short-axis views, while only one to three slices were acquired for long-axis views. $^{**}$Reported scan times do not include resting periods between breath-holds.}
\label{tab:params}
\end{table}

\subsection{Training}
The 22 volunteers are divided into three cohorts which comprise the training, validation, and test datasets (10, 2, 10 split). For training, each 4-D dataset (kx, ky, slice, cardiac phase) is split up slice-by-slice to create 180 unique 3-D training examples (kx, ky, cardiac phase). This number is further augmented using data augmentation techniques such as random flipping along readout and phase encoding directions with 50\% probability. Random circular translations were applied along the phase encoding direction by -20 to 20 pixels and along the cardiac dimension by -4 to +4 phases. Additionally, images are randomly center cropped along the readout direction to 64 points to reduce memory requirements for training. To show the network an adequate number of examples with anatomy overlap, the FOV in the phase encoding direction is retrospectively reduced by random factors between 0-15\%. Variable-density undersampling masks are randomly generated and applied on-the-fly to each example during training. These masks are designed to simulate 10-15X acceleration across the entire k-t plane. To ensure that the samples are well distributed, the acceleration rate is kept the same for each individual time point. Additionally, the first 20-30\% of k-space readout is masked out to simulate a partial echo acquisition. To simulate different temporal resolutions, the number of cardiac phases in the training data is varied between 12-20 by retrospectively sorting the data using a nearest neighbor gating approach. As a final pre-processing step, the undersampled complex-valued k-space data is scaled such that the pixel values in each coil-combined magnitude image lie in the range [0, 1].


To evaluate the impact of adding the extended coil sensitivity model, we trained two separate DL-ESPIRiT networks based on 3D convolutions with one set of maps (3DM1) and two sets of maps (3DM2). Additionally, to evaluate performance across 3D and (2+1)D networks, we trained and compared two other networks based on (2+1)D convolutions with one set of maps ((2+1)DM1) and two sets of maps ((2+1)DM2). All networks are comprised of 10 PGD iterations, one 5-layer ResNet per PGD iteration, and 96 feature maps per convolutional layer. As stated previously, the number of spatial feature maps in each (2+1)D convolution is expanded to 216 so that 3D and (2+1)D networks have the same number of learnable parameters (see Eq.~\ref{eq:separable}). In total, both networks contain 5,084,160 learnable parameters.

Both training and inference pipelines are implemented in TensorFlow \cite{Abadi2016}. In this implementation, the gradients of the loss function with respect to the network parameters are automatically computed in TensorFlow by backpropagating through the unrolled PGD network graph. Due to memory limitations, each network is trained with a batch size of 1 using the Adam optimizer \cite{Kingma2015} with hyperparameters $\beta_1$=0.9, $\beta_2$=0.999, $\epsilon$=10$^{-8}$, initial learning rate of 10$^{-3}$, and 200k training steps. A warm restart is performed after 100k training steps with a decayed learning rate of 10$^{-4}$ \cite{Loshchilov2016,Tran2018}. 3D and (2+1)D networks were trained for a total of 178 and 317 hours respectively.  To enable training a larger network with more PGD iterations, each DL-ESPIRiT network is split in half and trained across two NVIDIA Tesla V100 16GB video cards. However, note that training the network in this way incurs significant time overhead since data transfers between GPUs are necessary for each training step.

\subsection{Evaluation}
The performance of DL-ESPIRiT is evaluated on fully sampled data from which we can obtain ground-truth images. Specifically, ten fully-sampled short-axis cardiac cine datasets, excluded from the training, are retrospectively undersampled and reconstructed slice-by-slice using various reconstruction algorithms. We compare all 3D and (2+1)D DL-ESPIRiT approaches against a standard PICS reconstruction method known as $l_1$-ESPIRiT with spatial and temporal total variation (TV) regularization \cite{Uecker2014}. Regularization strengths for spatial and temporal TV priors are empirically determined to be 0.002 and 0.01 respectively, based on our pre-tuning of the two regularization parameters. The $l_1$-ESPIRiT problem is solved using the alternating direction method of multipliers \cite{Boyd2010} algorithm with 200 inner-loop iterations. We use the Berkeley Advanced Reconstruction Toolbox (BART, v0.4.04) implementation with GPU acceleration \cite{Uecker2015}. All reconstructions are performed on a separate computer system from the one used for training with one NVIDIA GTX 1080 Ti 12GB video card. 

To evaluate image quality for each reconstruction, peak signal-to-noise ratio (PSNR) and structural similarity index (SSIM) \cite{Wang2004} are computed for all slices and all volunteers in the test dataset. Both PSNR and SSIM metrics are evaluated with respect to corresponding fully-sampled references images inside of a bounding box drawn tightly around the heart of each subject. A two-tailed, paired $t$-test is conducted to determine the statistical significance of the improved image quality metrics between the different reconstruction methods. $l_1$-ESPIRiT and DL-ESPIRiT reconstructions are further evaluated based on the measurement accuracy of multiple cardiac functional indices including left ventricular end-diastolic volume (EDV), end-systolic volume (ESV), stroke volume (SV), and ejection fraction (LVEF). These indices are measured from automatic segmentations of epicardial and endocardial left ventricular borders extracted by Arterys Cardio DL (Arterys Inc, San Francisco, CA), a cloud-based medical image analysis tool which has been FDA cleared for LVEF measurement.  The Arterys segmentation is performed using a pre-trained convolutional neural network trained using a separate dataset comprised of 1,146 short-axis cine bSSFP exams annotated by board-certified radiologists \cite{Lieman-Sifry2017}. LVEF measurement accuracy is evaluated for each reconstruction with respect to automatic segmentations computed from fully-sampled images.

\section{Results}
\begin{figure}
\centering
\includegraphics[width=0.8\textwidth]{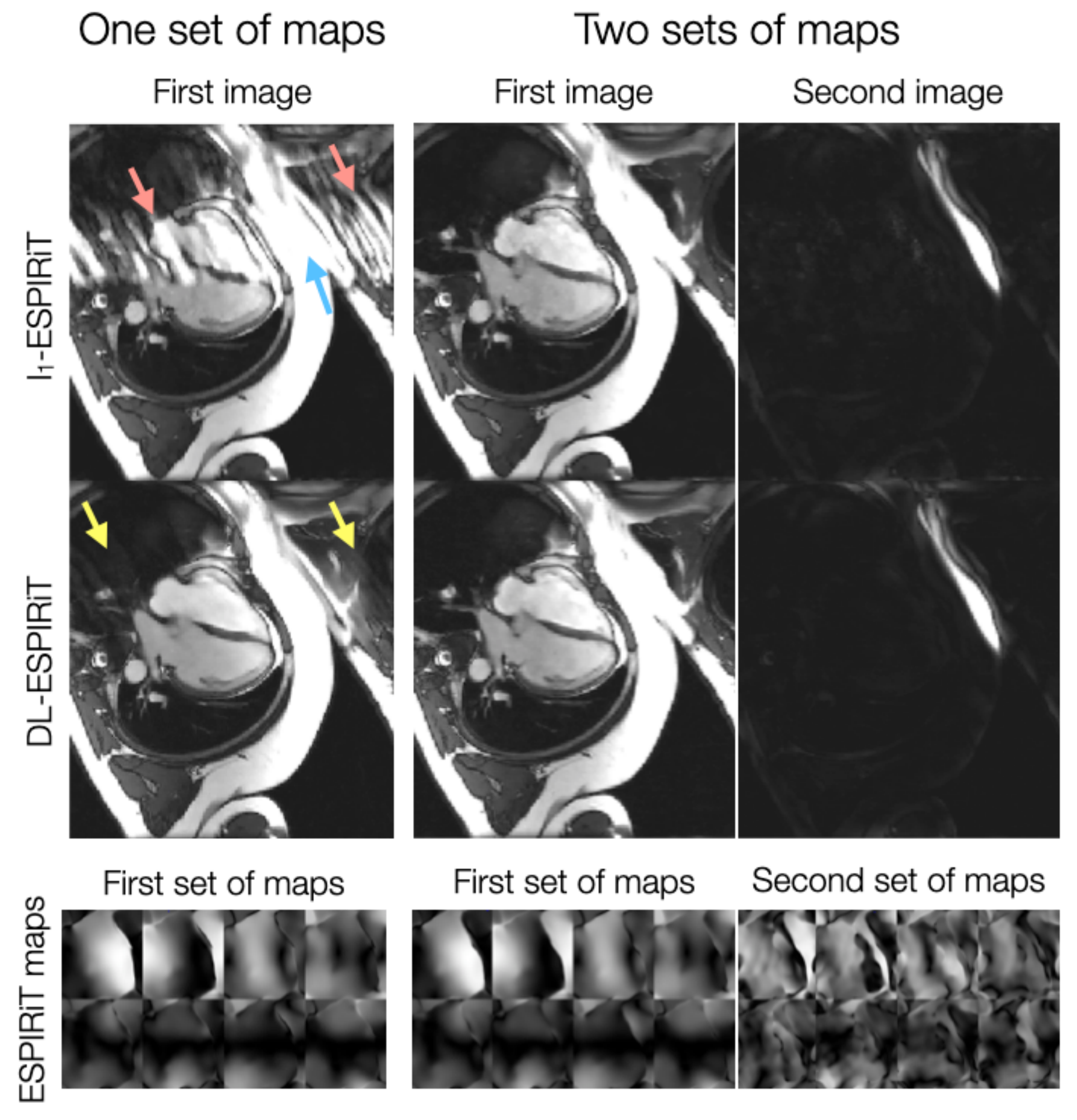}
\caption{A fully-sampled 2D cardiac cine acquisition in the standard 4-chamber view is retrospectively undersampled to simulate 10X acceleration, and then reconstructed using $l_1$-ESPIRiT and (2+1)D DL-ESPIRiT with one and two sets of ESPIRiT maps. Anatomy overlap is present along the chest wall (blue arrow), which causes significant ghosting along the phase encoding direction of the single-set $l_1$-ESPIRiT reconstruction (red arrows). Some ghosts are present in the single-set (2+1)D DL-ESPIRiT reconsruction (yellow arrows), but they are largely reduced compared to the single-set $l_1$-ESPIRiT reconstruction. Both double-set reconstructions are able to capture overlapping anatomies, separate them into two complex-valued channels (one for each set of maps), and completely eliminate ghosting artifacts. A corresponding video of these images is shown in Supporting Information Video S1.}
\label{fig:espirit}
\end{figure}

\begin{figure}
\centering
\includegraphics[width=1.0\textwidth]{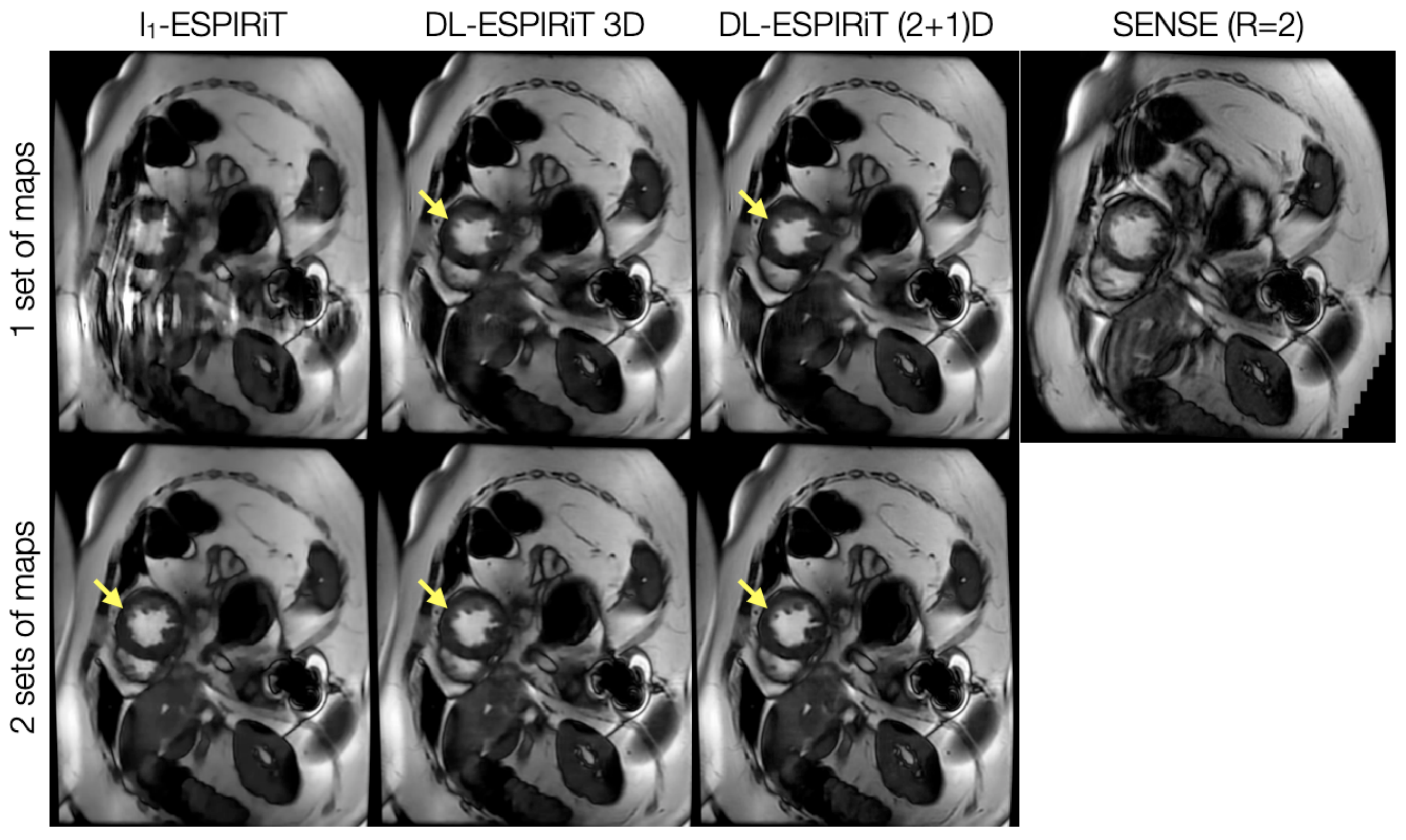}
\caption{With IRB approval, a prospectively undersampled (R=12) dataset is acquired from a pediatric patient within a single breath-hold on a 1.5T scanner. The images shown are reconstructed using $l_1$-ESPIRiT, 3D DL-ESPIRiT, and (2+1)D DL-ESPIRiT algorithms with one and two sets of ESPIRiT maps. For reference, a standard cardiac cine image acquired with 2-fold uniform undersampling (R=2) and reconstructed using SENSE is shown in the top right. Subtle motion artifacts arising from respiratory motion are shown in this image because the patient had difficulty holding their breath. Anatomy overlap is present along the chest wall, which causes severe aliasing on top of the heart as shown in the $l_1$-ESPIRiT reconstruction with one set of sensitivity maps. This artifact manifests as blurring in DL-ESPIRiT reconstructions with one set of ESPIRiT maps, and is most evident along left ventricle papillary muscles (yellow arrow). Both $l_1$-ESPIRiT and DL-ESPIRiT reconstructions resolve these artifacts when two sets of ESPIRiT maps are used. Corresponding videos for each reconstruction are shown in Supporting Information Video S3.}
\label{fig:results3}
\end{figure}

As shown by Fig.~\ref{fig:espirit}, DL-ESPIRiT reconstructions with one set of ESPIRiT maps show residual ghosting artifacts originating from sensitivity map errors in areas with anatomy overlap. By augmenting the signal model with two sets of ESPIRiT maps, the DL-ESPIRiT reconstruction is able to reconstruct overlapped components separately without ghosting artifacts. A corresponding video is shown in Supporting Information Video S1. An additional video of a 2-chamber view with anatomy overlap is shown in Supporting Information Video S2. This result is further demonstrated in a prospectively undersampled dataset shown in Figure \ref{fig:results3} with a corresponding video shown in Supporting Information Video S3.

\begin{figure}
\centering
\includegraphics[width=1.0\textwidth]{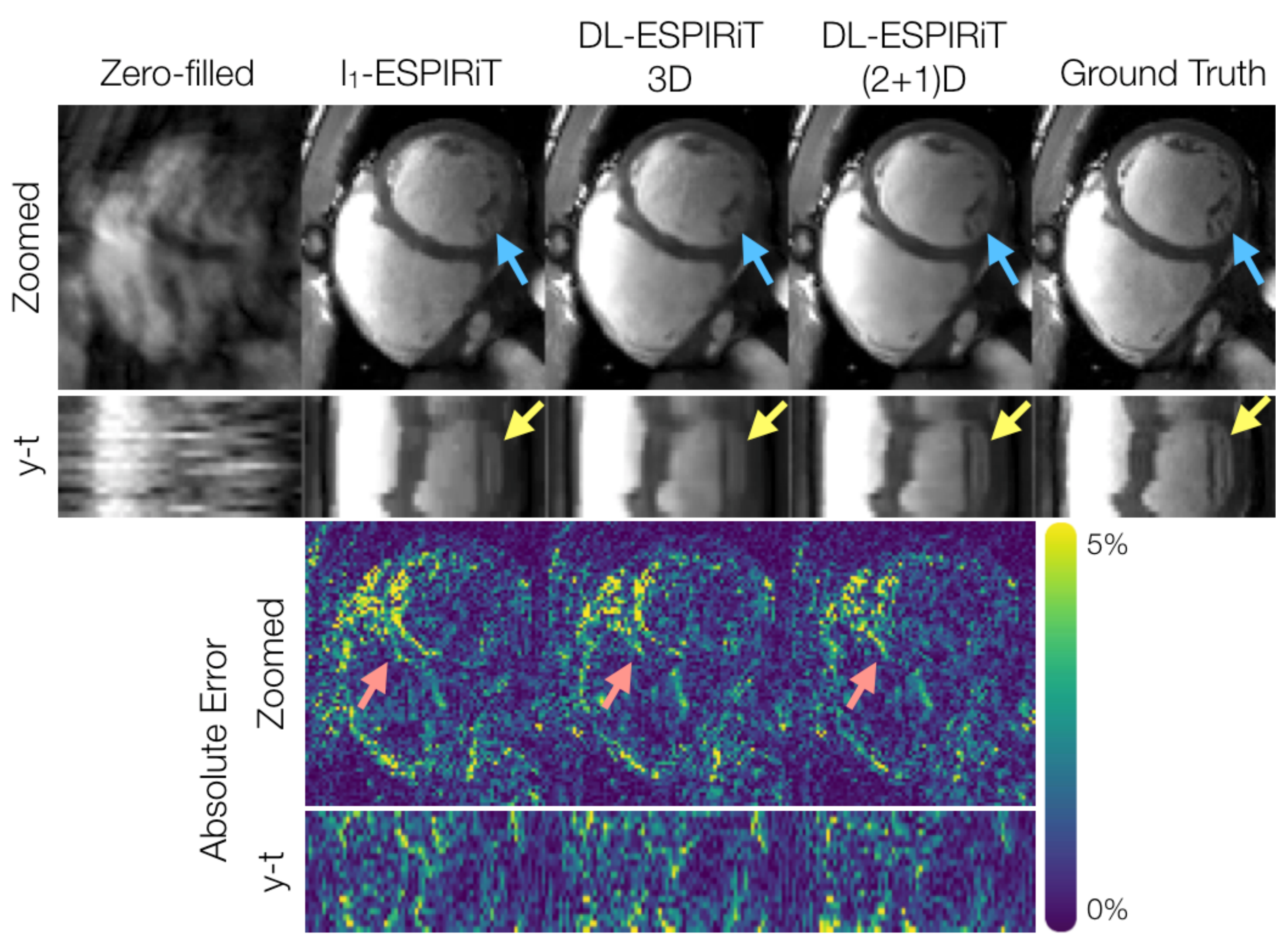}
\caption{A fully-sampled dataset acquired on a 1.5T scanner is retrospectively undersampled by a factor of 12, and reconstructed from left to right by: zero-filling, $l_1$-ESPIRiT, 3DM2 DL-ESPIRiT, and (2+1)DM2 DL-ESPIRiT methods. Only the first image corresponding to the first set of ESPIRiT maps are shown. Additionally, y-t profiles, error maps, and y-t error profiles are shown for each reconstruction. Both DL-ESPIRiT reconstructed images show reduced error around both epicardial and endocardial LV walls during mid-systole (red arrows). However, subtle movements in papillary muscle structures are not well depicted in either $l_1$-ESPIRiT or DL-ESPIRiT reconstructions (yellow arrows). Corresponding videos for magnitude and error images are shown in Supporting Information Video S4.}
\label{fig:results1}
\end{figure}

\begin{figure}
\centering
\includegraphics[width=0.9\textwidth]{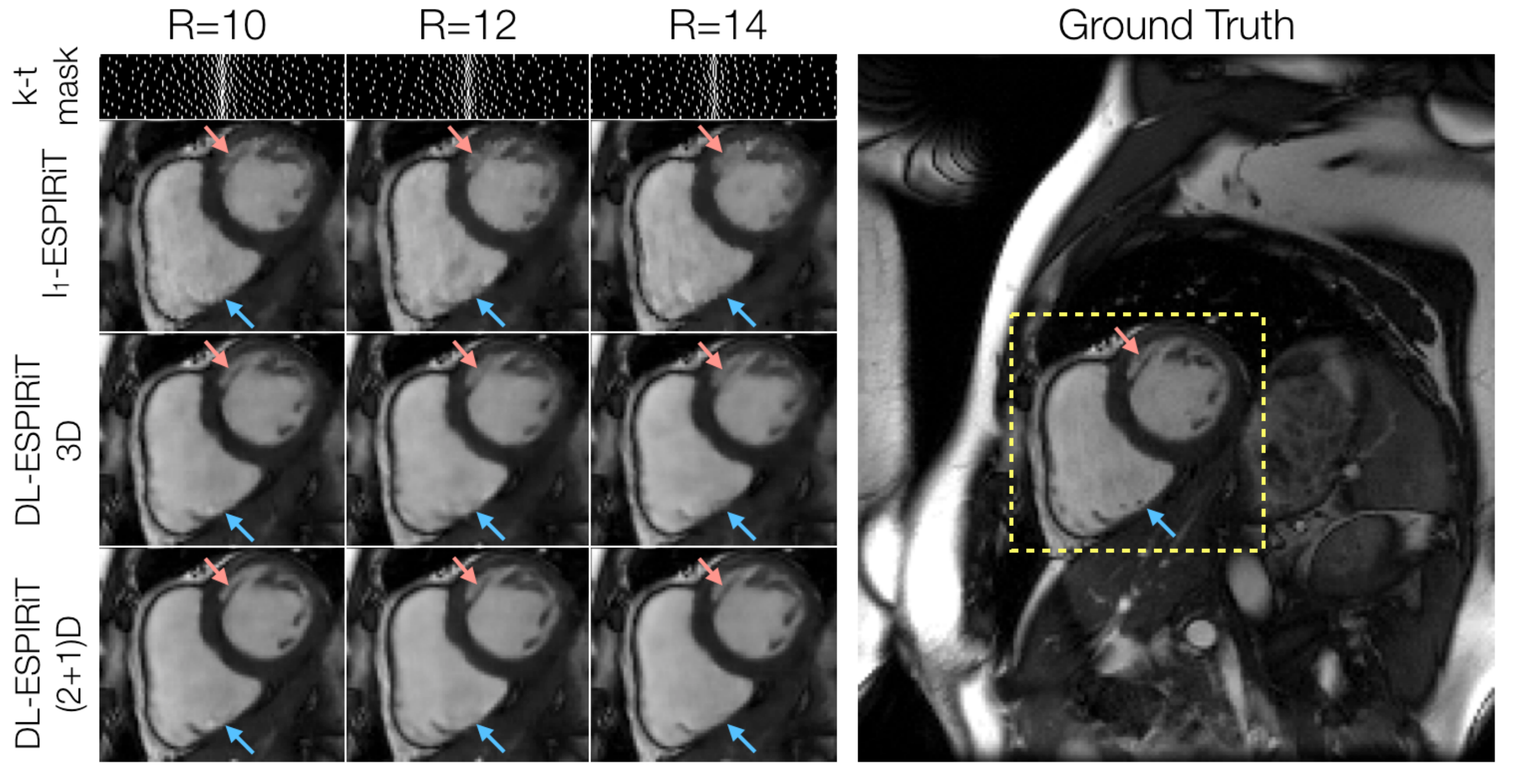}
\caption{A fully-sampled dataset acquired on a 3.0T scanner is retrospectively undersampled by factors of 10, 12, and 14 using variable density masks.  As expected, reconstructed mid-systolic frames become progressively blurrier as the acceleration rate is increased. This is particularly evident in smaller structures such as left ventricular trabeculations (red arrow). However, the (2+1)DM2 DL-ESPIRiT network retains sharpness of this structure as the acceleration rate is increased. However, none of the reconstruction methods were able to recover tiny papillary muscle structures inside of the right ventricle for this range of acceleration rates (blue arrows). Corresponding videos of reconstructions at each acceleration rate shown here are shown in Supporting Information Video S5.}
\label{fig:results2}
\end{figure}

Figure \ref{fig:results1} shows representative $l_1$-ESPIRiT, 3DM2 DL-ESPIRiT, and (2+1)DM2 DL-ESPIRiT reconstructions of data that was retrospectively undersampled by 12X. For this set of images, (2+1)DM2 DL-ESPIRiT is able to capture more realistic cardiac dynamics than $l_1$-ESPIRiT. $l_1$-ESPIRiT reconstructions show significant staircasing artifacts along time, which are characteristic of total variation-based reconstructions. Furthermore, error maps in Fig.~\ref{fig:results1} show that the (2+1)DM2 network produces higher fidelity images than both $l_1$-ESPIRiT and 3DM2 DL-ESPIRiT reconstructions. Corresponding videos depicting magnitude and error images are shown in Supporting Information Video S4. In Figure ~\ref{fig:results2}, the performance across all reconstruction methods for acceleration rates 10, 12, and 14 is compared. As the acceleration rate is increased, each method produces progressively blurrier images, except for the (2+1)DM2 DL-ESPIRiT reconstruction method, which retains sharpness of left ventricular trabeculae. A corresponding video showing reconstructions at each acceleration rate is shown in Supporting Information Video S5.

\begin{figure}
\centering
\includegraphics[width=0.9\textwidth]{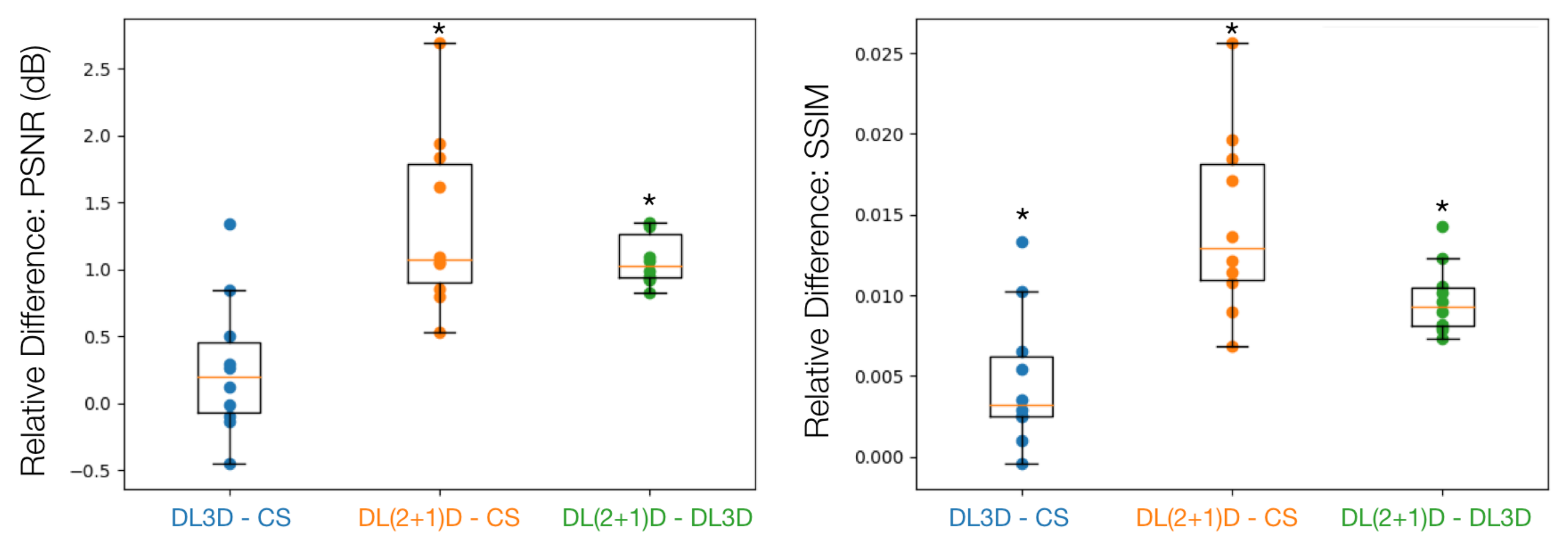}
\caption{Ten fully-sampled short-axis cine datasets are retrospectively undersampled by 12X and reconstructed using $l_1$-ESPIRiT (CS), 3D DL-ESPIRiT (DL3D), and (2+1)D DL-ESPIRiT (DL(2+1)D) all with two sets of ESPIRiT maps. PSNR and SSIM metrics are computed with respect to fully-sampled images within a manually drawn bounding box containing the heart tissue. The relative differences between image quality metrics for each reconstruction are plotted here. For example, if two reconstruction methods A and B are being compared, and the relative metric difference A - B is greater than zero, then method A performed better than method B. This box and whisker plot demonstrates that on average both DL methods performed better than CS, while DL(2+1)D performs better than DL3D with respect to PSNR and SSIM metrics. A * above the box plot indicates that the result is statistically significant ($p<0.01$) based on a paired $t$-test.}
\label{fig:metrics}
\end{figure}

Figure \ref{fig:metrics} shows the relative differences of PSNR and SSIM metrics between each reconstruction method for each subject in the test dataset. (2+1)DM2 DL-ESPIRiT significantly outperforms both $l_1$-ESPIRiT and 3DM2 DL-ESPIRiT with respect to PSNR and SSIM metrics ($p<0.01$). Additionally, the average reconstruction times for $l_1$-ESPIRiT, 3D DL-ESPIRiT, and (2+1)D DL-ESPIRiT were 5.36 $\pm$ 0.05, 3.89 $\pm$ 0.04, and 4.89 $\pm$ 0.03 seconds per 3-D (kx,ky,t) slice respectively.

\begin{figure}
\centering
\includegraphics[width=0.6\textwidth]{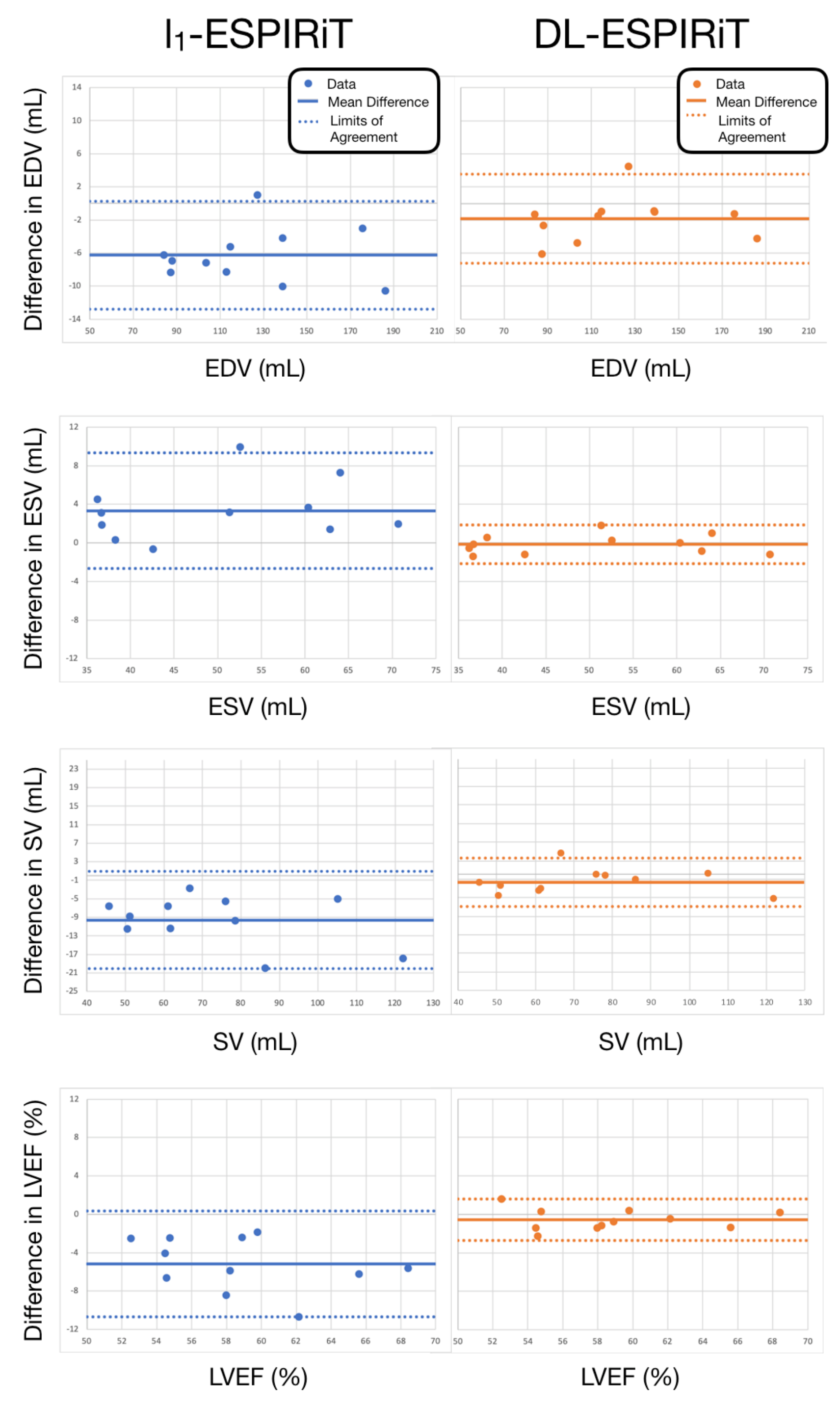}
\caption{Fully-sampled short-axis cine data are retrospectively undersampled by 12X and reconstructed using $l_1$-ESPIRiT, and (2+1)DM2 DL-ESPIRiT. Endocardial left ventricular borders are automatically segmented to compute end-diastolic volume (EDV), end-systolic volume (ESV), stroke volume (SV), and ejection fraction (LVEF) for each subject. Bland-altman plots show significantly better agreement between indices measured from DL-ESPIRiT reconstruction and ground truth measurements. The mean difference (solid lines) and limits of agreement (dashed lines) are reported here for each index as: mean, [lower limit, upper limit]. For $l_1$-ESPIRiT these are EDV: -6.27 [-12.79, 0.24] mL, ESV: 3.33 [-2.66, 9.32] mL, SV: -9.60 [-20.10, 0.89] mL, and LVEF: -5.15 [-10.67, 0.36] \%. For DL-ESPIRiT, these are EDV: -1.84 [-7.24, 3.56] mL, ESV: -0.15 [-2.16, 1.85] mL, SV: -1.68 [-7.00, 3.63] mL, and LVEF: -0.56 [-2.73, 1.60] \%.}
\label{fig:bland_altman}
\end{figure}

Figure \ref{fig:bland_altman} shows Bland-Altman plots for left ventricular EDV, ESV, SV, and EF measured from $l_1$-ESPIRiT and (2+1)DM2 DL-ESPIRiT reconstructed images. As a result of improved image quality and sharpness, automatic segmentations from DL-ESPIRiT images produce more accurate EDV, ESV, SV, and EF measurements than $l_1$-ESPIRiT with respect to measurements made from fully-sampled images.

\begin{figure}
\centering
\includegraphics[width=1.0\textwidth]{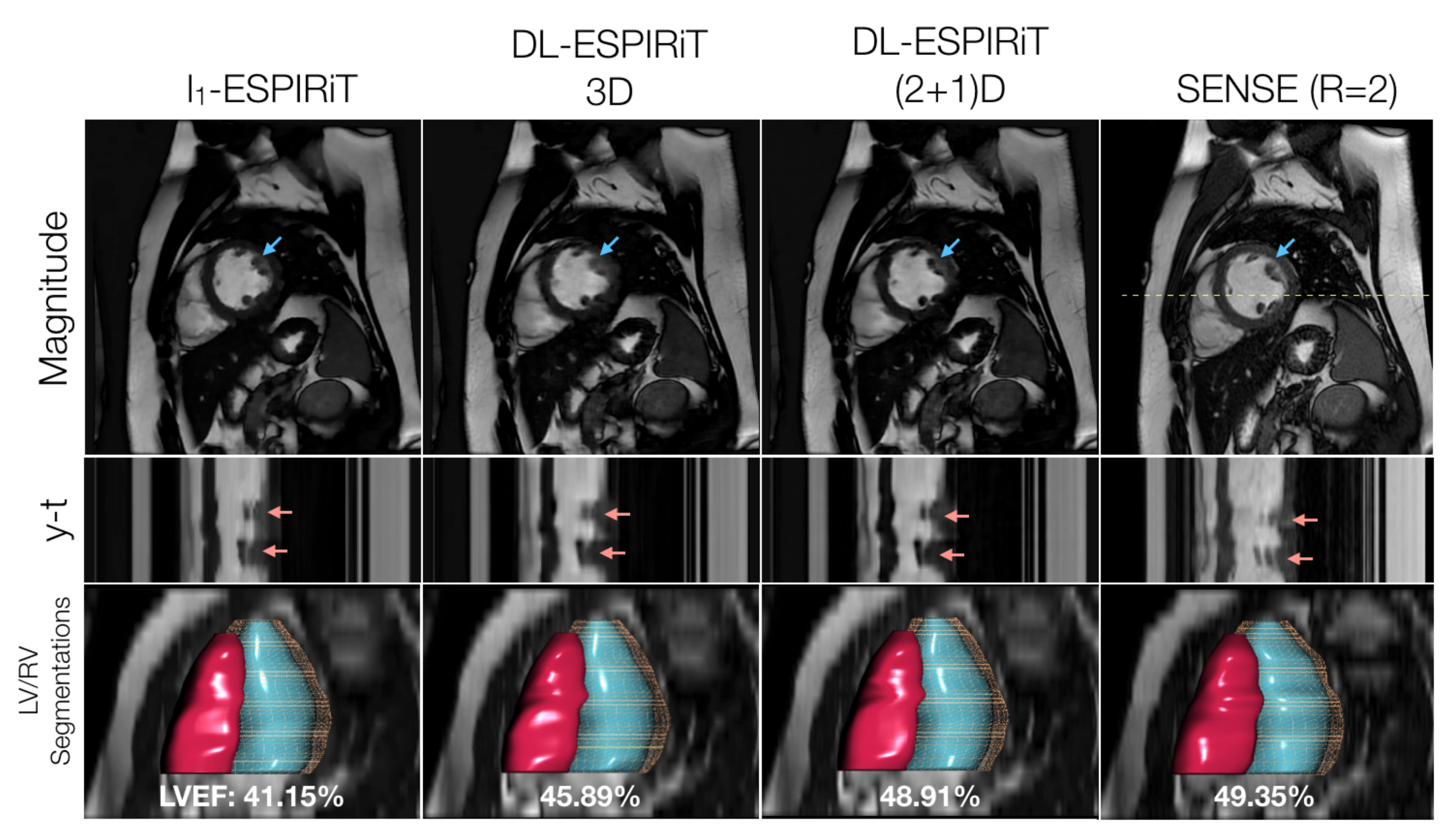}
\caption{With IRB approval, two prospectively undersampled 2D cardiac cine scans are performed in a pediatric patient with premature ventricular contractions (PVC), which manifest as an extra beat motion during the cardiac cycle. Performed as part of a routine cardiac MR exam, the first acquisition is uniformly undersampled (R=2) and reconstructed using SENSE. The second acquisition is undersampled using a variable-density k-t sampling pattern (R=12) and is reconstructed using $l_1$-ESPIRiT and DL-ESPIRiT algorithms. Mid-systolic frames, y-t profiles, and segmentations of the left and right ventricles are shown here. Despite the lack of subjects with PVC in the training data, the DL-ESPIRiT network is able to generalize to this particular example and reconstruct an accurate depiction of the extra beat motion that agrees with both $l_1$-ESPIRiT and SENSE reconstructions (red arrows). Furthermore, left ventricular ejection fraction (LVEF) is measured in each of the reconstructions, and it is shown that (2+1)DM2 DL-ESPIRiT is in better agreement with the clinical standard SENSE reconstruction than is $l_1$-ESPIRiT and 3DM2 DL-ESPIRiT. Corresponding videos for each reconstruction are shown in Supporting Information Video S6.}
\label{fig:results4}
\end{figure}

To demonstrate the generalizability of the proposed approach, Figure \ref{fig:results4} shows $l_1$-ESPIRiT and DL-ESPIRiT reconstructions of a prospectively undersampled dataset acquired in a patient with an abnormal heartbeat. Despite the lack of this type of cardiac motion in the training data, DL-ESPIRiT faithfully reconstructs the abnormal behavior when compared to the images from the clinical standard sequence. Corresponding videos of each reconstruction are shown in Supporting Information Video S6.

\section{Discussion}
In this work, a combined parallel imaging and deep learning-based reconstruction method known as DL-ESPIRiT is developed and trained to reconstruct vastly accelerated 2D cardiac cine MRI data. Through various analyses on reconstructions of retrospectively undersampled, our results suggest that (2+1)D DL-ESPIRiT may enable higher fidelity reconstructions than compressed sensing for single heartbeat acquisitions by leveraging historical exam data to learn better priors for constrained reconstruction. We build on previous data-driven reconstruction methods by incorporating a robust parallel imaging model, known as ESPIRiT, that is capable of leveraging multi-coil information to reconstruct images without SENSE-related FOV limitations. Additionally, a novel 3D CNN architecture based on separable 3D convolutions is proposed to simplify network training and achieve higher reconstruction accuracy for 2D cardiac cine data.

Within this framework, we propose a generalized method for reconstruction of multiple images corresponding to multiple sets of SENSE-like coil sensitivity maps derived using ESPIRiT. Compared to previously proposed frameworks that use a single set of sensitivity maps in their signal models \cite{Hammernik2018,Cheng2018}, DL-ESPIRiT uses multiple sets of ESPIRiT maps to enhance its robustness to errors in the map estimation process. This is especially advantageous for 2D cardiac cine imaging, which is performed in double oblique planes and thus prone to sensitivity map errors arising from anatomy overlap. Other data-driven learning methods that do not rely on sensitivity maps instead apply neural networks in k-space to directly synthesize missing data samples \cite{Han2019,Akcakaya2019}. These methods are also exempted from artifacts caused by erroneous sensitivity maps, however, they rely on local correlations in k-space to estimate missing data which may limit the choice of sampling pattern and increase computation compared to SENSE-like reconstructions. 

In this work, we compare our method against an existing PICS method known as $l_1$-ESPIRiT using two sets of sensitivity maps and a spatiotemporal total variation prior. As shown by reconstructions of 12X accelerated data in Figs.~\ref{fig:results1} and \ref{fig:results2}, $l_1$-ESPIRiT produces images with significant spatiotemporal blurring and staircasing artifacts. As a result, automatic segmentations made on $l_1$-ESPIRiT reconstructions produce less accurate measurements of ventricular volumes. In particular, end-systolic volumetric measurements are overestimated, which is likely caused by temporal blurring due to rapid cardiac motion near end-systole. This was also found in another study by Inoue et al, which found that 2D cardiac cine scans acquired with lower temporal resolution tend to produce overestimated ESV and underestimated LVEF \cite{Inoue2005}. 

On the other hand, DL-ESPIRiT produces visually sharper images and, as a result, more accurate LVEF measurements in reconstructions of retrospectively undersampled data. The DL-ESPIRiT network based on (2+1)D convolutions in particular produced the sharpest looking images with the best image quality on average. This can be attributed to multiple factors. Despite the 3D and (2+1)D networks having the same number of learnable parameters, the (2+1)D network exhibits better training convergence and reconstruction accuracy due to the convolutional kernels' simpler structure \cite{Tran2018}. Moreover, the additional ReLU activation layer in between spatial and temporal convolutions may enhance the (2+1)D network's representational power and therefore its ability to learn more complex priors than a conventional 3D network. Although higher representational power could cause the network to overfit the training data, this phenomenon is not observed in the monotonically decreasing validation loss curves shown in Fig.~\ref{fig:separable}C.

Despite improvements in image quality over $l_1$-ESPIRiT, DL-ESPIRiT reconstructions are notably blurred with high acceleration for both retrospectively and prospectively undersampled data. This is especially evident in fine structures, such as papillary muscles and trabeculae, which can appear significantly blurred during systolic phases. In Fig.~\ref{fig:results2} for example, both $l_1$-ESPIRiT and DL-ESPIRiT reconstructions remove small papillary structures in the right ventricle, which are present in the fully-sampled images. This remains a limitation of the proposed method. Sharpness could be improved by using different network architectures and training loss functions. In this work, we chose the ResNet architecture for the basis of each unrolled PGD step due to its simplicity and stable training properties. However, other network architectures may provide higher-quality reconstructions at the cost of higher training complexity \cite{Yaman2019}. Additionally, using complex convolutions and activation layers which can more naturally exploit structure in complex-valued images may further improve reconstruction quality \cite{Virtue2017,Trabelsi2018}. With respect to the training loss function, we chose to use an $l_1$ loss since it has been previously shown to produce visually sharper images than a standard $l_2$ loss \cite{Zhao2017,Ghodrati2019}. More recently, adversarial loss functions have been shown to outperform both $l_1$ and $l_2$ losses with respect to standard image quality metrics and blind radiologist scoring \cite{Mardani2019}. Exploring different training loss functions for enhancing sharpness of these structures will be the subject of future work. 

In previous works \cite{Hammernik2018}, deep learning-based reconstruction approaches were shown to be computationally much faster than compressed sensing-based ones. In this work, we found DL-ESPIRiT reconstruction times to be only slightly faster than $l_1$-ESPIRiT. The discrepancy between our work and previous work is due to 3D convolutions. While 3D convolutions provide a natural way of jointly exploiting spatial and temporal dimensions for reconstruction, they are computationally more complex and memory intensive when compared to 2D convolutions. (2+1)D convolutions are composed of 2D and 1D convolutions in series and, in theory, could be implemented using 2D convolutions instead of 3D convolutions as described in this work. However, 2D convolutions only accept 4D arrays in the $NHWC$ (batch size, height, width, channels) format. Therefore, inputting dynamic data would require looping over the time dimension for the 2D spatial convolution, and looping over a spatial dimension for the 1D temporal convolution. This implementation requires significant memory overhead during training, which ultimately limits network depth and performance.


One limitation of this study is that the training data was retrospectively undersampled to numerically simulate data collected with a faster scan time. This was done in order to provide a proper baseline for comparison and evaluation of $l_1$-ESPIRiT and DL-ESPIRiT reconstruction techniques. However, training on simulated data may be suboptimal since prospectively collected data may contain features which are not present in the training data. In bSSFP imaging, for example, fast transitions between phase encoding steps gives rise to eddy currents that can perturb the steady state signal and introduce image artifacts \cite{Bieri2005}. These were not observed in our preliminary data with prospective undersampling shown in Figures \ref{fig:results3} and \ref{fig:results4}, but may arise for datasets collected with higher acceleration rates, smaller fields-of-view, or finer in-plane spatial resolution. Different eddy current compensation strategies such as phase encode pairing \cite{Bieri2005} will be explored in the future.


\section{Conclusion}
A novel deep learning-based reconstruction framework known as DL-ESPIRiT is proposed which improves robustness to SENSE-related FOV limitations over existing deep learning frameworks. Furthermore, a novel CNN architecture based on (2+1)D convolutions is proposed and integrated into DL-ESPIRiT to enhance spatiotemporal learning for dynamic image reconstruction. As a result of these two developments, a rapid single heartbeat per slice 2D cardiac cine scan is feasible, which can potentially lead to more accurate ventricular function assessment and improved patient comfort.


\clearpage

\section*{List of Supporting Information Videos}
\begin{itemize}
    \item[S1] This video depicts reconstructions of 4-chamber view data that has been retrospectively undersampled by a factor of 10 (corresponding to images shown in Fig.~\ref{fig:espirit}). The top row depicts $l_1$-ESPIRiT, 3D DL-ESPIRiT, and (2+1)D DL-ESPIRiT reconstructions with one set of ESPIRiT maps. The bottom row depicts $l_1$-ESPIRiT, 3D DL-ESPIRiT, and (2+1)D DL-ESPIRiT reconstructions with two sets of ESPIRiT maps.
    \item[S2] This video depicts reconstructions of 2-chamber view data that has been retrospectively undersampled by a factor of 10. Similar to S1, the top row depicts $l_1$-ESPIRiT, 3D DL-ESPIRiT, and (2+1)D DL-ESPIRiT reconstructions with one set of ESPIRiT maps. The bottom row depicts $l_1$-ESPIRiT, 3D DL-ESPIRiT, and (2+1)D DL-ESPIRiT reconstructions with two sets of ESPIRiT maps.
    \item[S3] This video depicts reconstructions of short-axis view data that has been prospectively undersampled by a factor of 12 (corresponding to images shown in Fig.~\ref{fig:results3}). The top row depicts $l_1$-ESPIRiT, 3D DL-ESPIRiT, and (2+1)D DL-ESPIRiT reconstructions with one set of ESPIRiT maps. The bottom row depicts $l_1$-ESPIRiT, 3D DL-ESPIRiT, and (2+1)D DL-ESPIRiT reconstructions with two sets of ESPIRiT maps. The yellow arrow points to residual ghosting in the DL-ESPIRiT reconstructions caused by anatomy overlap. Additionally, a SENSE reconstruction of the 2-fold uniformly undersampled reference scan is shown in the top right corner of the video.
    \item[S4] This video depicts magnitude (top row) and absolute error maps (bottom row) of reconstructions that have been retrospectively undersampled by a factor of 12 (corresponding to the images shown in Fig.~\ref{fig:results1}). Data is reconstructed (from left to right) by: $l_1$-ESPIRiT, 3D DL-ESPIRiT, and (2+1)D DL-ESPIRiT algorithms with two sets of sensitivity maps. The absolute error video is computed by taking the absolute difference between each reconstruction and the ground truth for each frame.
    \item[S5] This video depicts reconstructions of short-axis view data that has been retrospectively undersampled by different factors (coresponding to images shown in Fig.~\ref{fig:results2}). Each column represents 10X, 12X, and 14X acceleration rates. Each row depicts $l_1$-ESPIRiT, 3D DL-ESPIRiT, and (2+1)D DL-ESPIRiT reconstructions with two sets of sensitivity maps. The bottom right video depicts the fully-sampled ground truth images.
    \item[S6] This video depicts reconstructions of data that has been prospectively undersampled by a factor of 12 (corresponding to images shown in Fig.~\ref{fig:results4}). Data is reconstructed (from left to right) by: $l_1$-ESPIRiT, 3D DL-ESPIRiT, and (2+1)D DL-ESPIRiT algorithms with two sets of sensitivity maps. For reference, the far right video is a SENSE reconstruction of 2-fold uniformly undersampled data which was acquired as part of a routine cardiac MRI exam.
\end{itemize}

\end{document}